\documentclass{aa}

\usepackage{graphicx}
\usepackage{color}
\usepackage[varg]{txfonts}
\usepackage{natbib}

\usepackage[table]{xcolor}

\begin{document}

\title {Categorization model of moving small-scale intensity enhancements in solar active regions}

\author{ B.M. Shergelashvili \inst{1,2,3}, E. Philishvili \inst{2,3}, S. Buitendag \inst{4}, S. Poedts \inst{5,6}, M.~Khodachenko\inst{1,7,8} }

\institute{
Space Research Institute, Austrian Academy of Sciences, Schmiedlstrasse 6, 8042 Graz, Austria\\
 \and
Centre for Computational Helio Studies, Ilia State University, G.\ Tsereteli street 3, 0162 Tbilisi, Georgia\\
  \and
Evgeni Kharadze Georgian National Astrophysical Observatory, M.\ Kostava street 47/57, 0179 Tbilisi, Georgia\\
 \and
  Capitec Bank, 5 Neutron Street, Stellenbosch, 7600, South Africa \\
 \and
  Centre for Mathematical Plasma Astrophysics/Department of Mathematics, Celestijnenlaan 200B, 3001, Leuven, Belgium\\
  \and
Institute of Physics, University of Maria Curie-Sk{\l}odowska, 20-031 Lublin, Poland\\
\and
Institute of Astronomy, Russian Academy of Sciences, 119017, Moscow, Russia\\
\and
Institute of Laser Physics SB RAS, 630090 Novosibirsk, Russia\\
}

\abstract
{The small-scale moving intensity enhancements remotely observed in the extreme ultraviolet images of the solar active regions, which we refer to as active region moving campfires (ARMCs), are related to local plasma temperature and/or density enhancements. Their dynamics is driven by the physical processes in the entire coronal plasma. Our previous study of ARMCs indicates that they have characteristic velocities at around the background sound speed. In the present paper, we further investigate the dynamical and statistical properties of ARMCs.}
{The main goal of our work is to carry out a simultaneous analysis of EUV images from two observational missions, SDO/AIA and Hi-C 2.1. The aims of the performed cross-validating analysis of both SDO/AIA and Hi-C 2.1 data were to reveal how the observed moving features are distributed over the studied active region, AR12712, and to perform a statistical hypothesis test of the existence of different groups of ARMCs with distinct physical characteristics.}
{We use the statistical model of intensity centroid convergence and tracking that was developed in our previous paper. Furthermore, a Gaussian mixture model fit of the observed complex of moving ARMCs is elaborated to reveal the existence of distinct ARMC groups and to study the physical characteristics of these different groups.}
{In data from the 171\AA, 193\AA\ and 211\AA\ channels of SDO/AIA, we identified several groups of ARMCs with respect to both blob intensity and velocity profiles. The existence of such groups is confirmed by the cross-validation of the 172\AA\ data sets from Hi-C 2.1.}
{The ARMCs studied in this paper have characteristic velocities in the range of the typical sound speeds in coronal loops. Hence, these moving objects differ from the well-known rapid Alfv\'enic velocity jets from magnetic reconnection sites. This is also proven by the fact that ARMCs propagate along the active region magnetic structure (strands). The nature of the discovered statistical grouping of the ARMC events is not known. Further theoretical studies and modeling is required to reveal this nature.}


\titlerunning{Categorization of small-scale intensity enhancement in solar active region}

\authorrunning{Shergelashvili, Philishvili, Buitendag et al.}

\maketitle

\section{Introduction}
The kinematics of solar coronal plasma motions characterizes the rate of energy deposition into the corona. Therefore, the observed moving small-scale brightenings in subsequent images of the solar corona have been of specific scientific interest.  These small features with a relatively small velocity amounting to 2-6$\;$km/s \citep{rueedi1992,degenhardt1993} are usually connected to local magnetic structures and the physical processes that occur in them. Theoretically, they are interpreted as siphon-type flows \citep{montesinos1993}.

The coordinated analysis of data sets from different observational missions is needed to infer reliable conclusions about these kinds of motions. In this context, high-resolution observations of the solar coronal active regions within the framework of the High Resolution Coronal Imager missions Hi-C \citep{HIC2014} and Hi-C 2.1 \citep{HIC22019} represent valuable sources of data. The recent measurements of properties of a large population of the so-called coronal strands in the active region AR~12712 observed by Hi-C 2.1 and the Atmospheric Imaging Assembly \citep[AIA;][]{AIA2012} instrument of the Solar Dynamics Observatory (SDO) are reported in \citet{WilliamsHIC22020}. In the present work, we reconsider the same active region. Our two major goals are: (1)~to perform a combined analysis of data from the Hi-C 2.1 and SDO/AIA missions, aiming at the discovery of the small-scale intensity enhancements (local extreme ultraviolet brightenings) and their motion properties in the hosting multi-strand magnetic structures, and (2)~to check whether these phenomena are grouped with respect to certain characteristics. These small-scale intensity enhancements are referred to here as active region moving campfires (ARMCs). To achieve our goals, we apply the mathematically consistent method of ARMC convergence and tracking to the data sets. This method was presented in a previous paper, \citet{Philishvili2021}. Next, we perform a cluster analysis of the identified ARMCs to expose distinct groups of ARMCs with similar physical characteristics.  

Generally speaking, there is a large variety of different small-scale dynamical phenomena observed in the solar corona. In particular, one specific type of such small-scale motions, usually referred to as nanojets, originated from the pioneering work of \citet{parker1988}, who introduced the concept of the solar corona as a swarm of tiny magnetic reconnection events or nanoflares. Their direct detection, however, is restricted by the insufficient resolution of modern observational technologies. Hence, the nanoscale intensity brightenings moving in the complex magnetic realm of the solar atmosphere with various velocities presumably represent some indirect evidence of such magnetic energy release events; they would be accompanied by tiny jet-like outflows from the reconnection sites that have approximately Alfv\'enic velocities. Observations of nanojets have recently become available \citep{Antolin2021} through coordinated observations with the SDO/AIA instrument, the Interface Region Imaging Spectrograph \citep{IRIS2014}, and the Hinode/Solar Optical Telescope \citep{HINODE12008,HINODE22008}. The main characteristics of such observed motions can be interpreted as the motions of plasma at Alfv\'en speed across the magnetic field of coronal loops at the sites of small angle magnetic reconnection \citep{Antolin2021}. From the point of view of statistical mechanics, the occurrence of small reconnection events can be characterized by the stochastic process of the nonequilibrium diffusion of magnetic features \citep[for theoretical aspects, see, e.g.,][]{maesbook2009,Maes2009}. It is commonly accepted that such a system can achieve the statistical state of self-organized criticality, resulting in an avalanche-type spatiotemporal behavior \citep{Antolin2021,SOCAschwanden12016,Baiesi2008}. 

Another type of the observed motions are the rapid plasma jet outflows \citep[with characteristic speeds of up to 100-200 km/s;][]{anfinogentov2021} from an active region \citep{Joshi2020} and from coronal hole bright points \citep{Bagashvili2018} that manifest the oscillatory behavior of their precursors. A catalog of hot jet events has also been reported \citep{Kaltman2021}. Recently, in preliminary imaging data sets of the Extreme Ultraviolet Imager \citep{Rochus2020} from the Solar Orbiter mission, small-scale brightenings (so-called campfires) have also been observed in the quiet Sun \citep{bergmans2021}.  

The subject of the present study is related to a special kind of extreme ultraviolet (EUV) intensity brightenings in the active regions, namely the so-called ARMCs.  \citet{Philishvili2021} proposed a method for the automatic identification and tracking of plasma flows in active regions and showed that the characteristic velocity of ARMCs ranges from sub- to supersonic values. As mentioned above, the methodology of \citet{Philishvili2021} was applied to EUV imaging data of active region AR 12712, which was observed on 29 May 2018. Altogether, 78 images of  AR 12712 were captured by the Hi-C2.1 at 172\AA\ between 18:56 and 19:02 UT with a cadence of 4.4~s and a plate scale of 0.$''$ 129 pixel$^{-1}$ \citep{HIC22019}. The Hi-C2.1 has almost the same EUV instrument as that on SDO/AIA 171\AA;\ 12~s; 1.$''$2 pixel$^{-1}$ \citep{AIA2012}. Therefore, in the present study we use the co-aligned images from both missions in multiple wavelengths for the same time window. 

The paper is structured as follows. In Sect.~2 details on the analysis methodology and its application to the considered data from SDO/AIA and Hi-C2.1  are provided. In Sect.~3 we discuss the obtained results. Section~4 contains our conclusions and outlines the perspectives of a possible follow-up study that would include further applications of the applied analysis methodology.

\section{Data sets and method of analysis}

The methodology contains four main parts or steps. First, we reveal the physical characteristics of ARMCs observed in SDO/AIA images, and we pay particular attention to differences in these characteristics that may indicate the presence of distinct groups.
Secondly, we determine the number of groups with respect to each characteristic by means of statistical analysis and hypothesis testing. Thirdly, we apply the same approach to the Hi-C2.1 data.
Finally, we illustrate the existence and location of each characteristic group and briefly discuss the similarities between SDO/AIA and Hi-C2.1 pictures. 

\subsection{Characteristics and grouping of ARMCs, according to SDO/AIA data}

\begin{figure}[ht!]
\centering
\includegraphics[width=80mm]{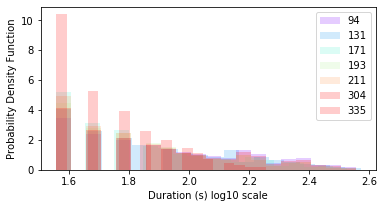}
\includegraphics[width=80mm]{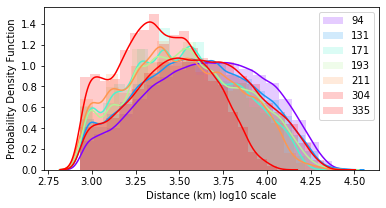}
\includegraphics[width=80mm]{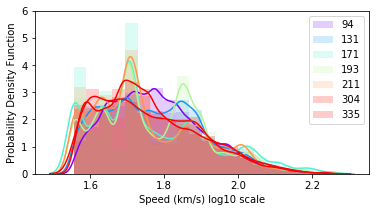}
\includegraphics[width=80mm]{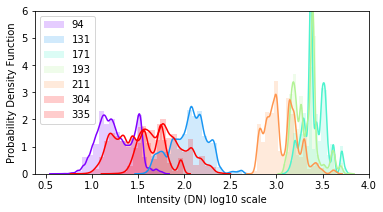}
\caption{{Distribution of the ARMC characteristics in terms of  histograms (shaded vertical bars) and kernel densities (solid lines), revealed from the SDO/AIA images for the different wavelengths. From top to bottom: Duration, distance, speed, and intensity. The duration and distance distributions are both unimodal, which indicates that they all belong to one single group. The speed and intensity distributions, however, are multimodal, which indicates the existence of distinct groups of ARMCs.}}
\label{hist}
\end{figure}

We started by considering four characteristics of the moving ARMCs, namely their {duration ($T_k$), trajectory distance ($D_k$), speed ($V_k$), and intensity enhancement ($I_k$), which are derived from the tracked flow coordinates $(x_k,y_k)$ at time $t_k$ for each respective frame, $k$, as follows:}
\begin{equation}
    T_{k} = \sum_{i=2}^k \, t_i-t_{i-1}, 
\end{equation}
\begin{equation}
    D_{k} = \sum_{i=2}^k \, \sqrt{(x_i-x_{i-1})^2+(y_i-y_{i-1})^2}, 
\end{equation}
\begin{equation}
    V_{k} = \frac{D_{k}-D_{{k-1}}}{t_k-t_{k-1}},
\end{equation}
\begin{equation}
    I_k = I(x_k, y_k), 
\end{equation}
{where $I(x_k,y_k)$ denotes the pixel intensity of frame $k$ at the center of the brightening $(x_k,y_k)$.}
The distributions of all these characteristics are illustrated in Fig.~\ref{hist} by {both} a histogram and a kernel density plot for each of the seven different SDO/AIA wavelengths. This provides visual insights  regarding the existence of groups of flows with similar characteristics. The duration and distance distributions are both unimodal, which implies that each of them contains no more than one group. 

\begin{figure}[ht!]
\centering
\includegraphics[width=80mm]{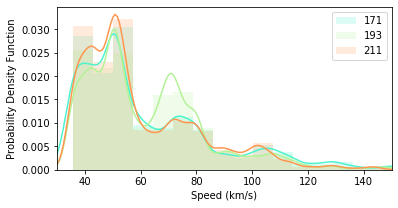}
\includegraphics[width=80mm]{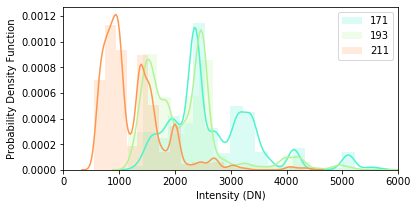}
\caption{Distribution of the ARMC speed and intensity in 171\AA, 193\AA\ and 211\AA\. The speed distribution is similar across all three wavelengths. The intensity distributions differ due to their explicit dependence on each respective wavelength but are similar where they overlap.}
\label{density}
\end{figure}

The speed and intensity distributions, however, are multimodal for certain wavelengths. In particular, the speed distribution is multimodal for the 171\AA, 193\AA\ and 211\AA\ wavelengths, whereas the intensity distribution is multimodal for all wavelengths. This provides evidence that there indeed exist several groups with different speed and intensity characteristics. Further on we consider only the wavelengths 171\AA, 193\AA\ and 211\AA\ to study ARMCs observed in SDO/AIA images. The reason for such a restriction is that only for these wavelengths are the hints of grouping in the speed and intensity distributions seen simultaneously (e.g.,\ in Figs.~\ref{hist} and \ref{density}). 

In order to better understand the speed and intensity characteristics of the ARMC groups, the Gaussian mixture models (GMMs) were fitted to the respective data sets using the expectation-maximization algorithm \citep{Dempster1977}. This yielded the proportion, location, and scale estimates for each group. The proportion estimate indicates what percentage of all flows belongs to a given group, the location estimate is a measure of the expected speed or intensity of the particular ARMC group, and the scale parameter characterizes the variation of the speed or intensity in a group around its expected value. In fact, the ARMC grouping was judged on the bases of all these estimates. A flow belongs to the group with which {it} has the largest ``responsibility.'' This responsibility is a relative density function value for the given flow's intensity or speed, as defined in \citet{Hastie} and \citet{Hastie2}. 

When fitting the GMMs, determining the optimal number of the groups is the most important factor. It has to strike a balance between a good distribution fit (the more groups, the better the fit) and whether the groups are distinguishable (the fewer groups, the easier it is to distinguish one group from another). 
{The number of groups was chosen as the number of distinct local maxima of the density function, which is illustrated in Fig.~\ref{density}. This number is four and six for speed and intensity, respectively. This conclusion was assessed using two approaches. Firstly, the Bayes information criterion (BIC) and the Akaike's information criterion (AIC) were plotted to determine the goodness-of-fit \citep{Hastie}.  Secondly, statistical hypothesis tests were performed, specifically the Anderson-Darling hypothesis test and the Cramer-von Mises hypothesis test.}

\begin{figure}[ht!]
\centering
\includegraphics[width=80mm]{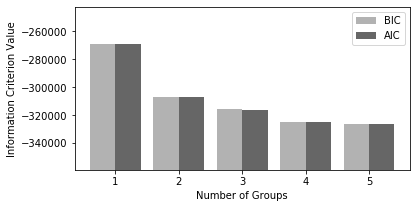}
\includegraphics[width=80mm]{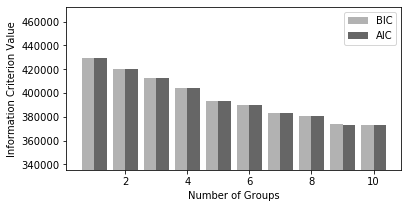}
\caption{ BIC and AIC of the GMM fit to the combined 171\AA, 193\AA\ and 211\AA\ wavelength flow speed (top) and intensity (bottom).}
\label{bic_aic}
\end{figure}

Figure~\ref{bic_aic} clearly shows that both the BIC and AIC values approximately saturate at four and six groups for the ARMC speeds and intensities, respectively, indicating that this amount of  groups has an adequate goodness-of-fit while being clearly distinguishable. The Anderson-Darling and Cramer-von Mises hypothesis tests assess how close the fitted GMM's cumulative distribution function ($\hat{F}$) is to the empirical cumulative distribution function ($F_n$) of the considered sample of ARMCs. The null hypothesis for both tests is that the observed sample is generated from the fitted GMM. The degree of deviation from this hypothesis statement is characterized by a p-value.
The higher the p-value for these tests, the lower the confidence level at which the null hypothesis is rejected. A higher p-value therefore indicates a better fit. {The statistics of both hypothesis tests can be written as follows:}
\begin{equation}
T_w = n \, \int_{-\infty}^{\infty} \, w(u) \, \left( F_n(u) - \hat{F}(u) \right)^2 \, d\hat{F}(u),
\end{equation}
{where $w(u)=1$ for the Cramer-von Mises test statistic and $w(u)=\left(\hat{F}(u)\,(1-\hat{F}(u))\right)^{-1}$ for the Anderson-Darling test statistic, and $n$ is the sample size. This test statistics are effectively a weighted mean of the squared difference between the fitted GMM distribution function, $\hat{F}(u)$, and the empirical distribution function, $F_n(u)$, where the variable $u$ can be interpreted as the characteristic being tested, in this case either speed or intensity. The larger the difference between $\hat{F}$ and $F_n$, the larger the test statistic value, which means a smaller p-value and, therefore, a higher level of confidence with which the null hypothesis (that the observed sample is generated from the fitted GMM) is rejected. The difference between the two tests is that the Anderson-Darling hypothesis test places more weight on the tails of the fitted GMM than the Cramer-von Mises hypothesis test.
If the p-value of the Anderson-Darling hypothesis test is higher than the p-value of the Cramer-von Mises hypothesis test, then the GMM fits the tail of the distribution more closely than the body.}
The results of these hypothesis tests are summarized in Table~\ref{speedtests} for the flow speeds and in Table~\ref{intensitytests} for the flow intensities.  

\begin{center}
\begin{table}[ht]
\begin{tabular}{c|c|c}
\hline \hline
    & Anderson-Darling & Cramer-von Mises \\
Number of groups        &  test statistic p-value &  test statistic p-value \\
\hline \hline 
1   & 0.00060  & $<$0.00001 \\
2   & 0.06176    & 0.19117 \\
3   & 0.03964    & 0.35227 \\
4   & 0.36096      & 0.45189 \\
\hline \hline
\end{tabular}
\caption{Statistical hypothesis tests for the GMM goodness-of-fit regarding the speeds of ARMCs observed with SDO/AIA.}
\label{speedtests}
\end{table}
\end{center} 

The GMMs that consist of one, two, or three groups all have corresponding Anderson-Darling hypothesis test p-values under 10\%. We therefore reject the hypothesis that the observed flow speeds are generated from a Gaussian distribution with three or fewer groups at a 90\% level of significance. 
The p-value corresponding to both hypothesis tests that the observed flow speeds are generated from a GMM of four groups is greater than 35\%. We cannot, therefore, reject the hypothesis that the observed flow speeds are generated from a GMM of four groups at a 90\% level of significance, or even a 65\% level of significance. 
\begin{center}
\begin{table}[ht]
\begin{tabular}{c|c|c}
\hline \hline
    & Anderson-Darling & Cramer-von Mises \\
Number of groups        &  test statistic p-value &  test statistic p-value \\
\hline \hline 
1   & 0.00060  & 0.00018 \\
2   & 0.00158   & 0.00977 \\
3   & 0.02204    & 0.09553 \\
4   & 0.11598     & 0.11206 \\
5   & 0.10576    & 0.09993 \\
6   & 0.11967     & 0.12357 \\
\hline \hline
\end{tabular}
\caption{Statistical hypothesis tests for the GMM goodness-of-fit regarding the intensities of ARMCs observed with SDO/AIA.}
\label{intensitytests}
\end{table}
\end{center} 

\begin{figure*}
\centering
\includegraphics[width=80mm]{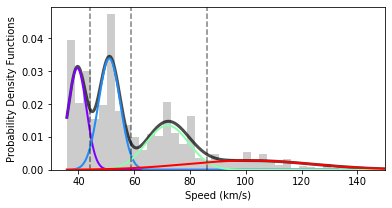}
\includegraphics[width=80mm]{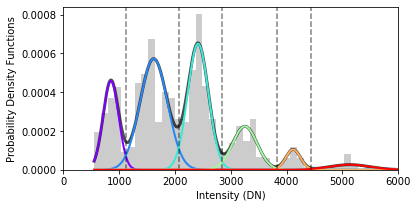}
\caption{GMM fit to the combined 171\AA, 193\AA\ and 211\AA\ wavelength flow speed (top) and intensity (bottom). The vertical lines represent the points where one Gaussian density crosses another, which indicates a separation of the ARMC groups.}
\label{gaussian_mix}
\end{figure*}

The p-values corresponding to the Cramer-von Mises hypothesis tests that the flow intensities are generated from a GMM consisting of five groups is less than 10\%. This means that we can reject the hypothesis that the observed flow speeds are generated from a Gaussian distribution with five groups at a 90\% level of significance. 
The fitted GMM of six groups has hypothesis test p-values larger than 10\%. Therefore, we cannot reject the hypothesis that the observed flow intensities are generated from a GMM of six groups at a 90\% level of significance.  \\
The resulting parameter estimates are used to group flows according to their respective responsibilities as discussed above. Figure~\ref{gaussian_mix} illustrates the Gaussian density functions of each ARMC group (colored lines) together with the resulting GMM's density function (black line). The vertical dashed lines indicate the group splits, which is the point at which a flow's responsibility to both groups is equal.

\begin{figure*}
\centering
\includegraphics[width=45mm]{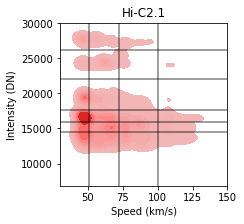}
\includegraphics[width=45mm]{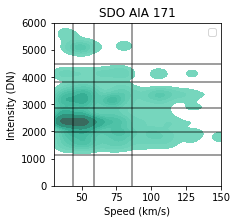}
\includegraphics[width=45mm]{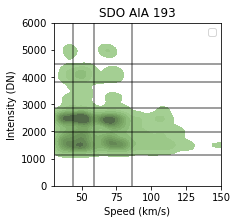}
\includegraphics[width=45mm]{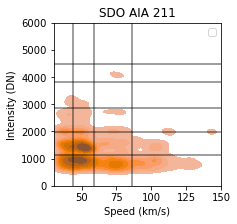}
\caption{ARMC grouping in the intensity-speed maps revealed from Hi-C2.1 172\AA\  data (left panel) and the SDO/AIA 171\AA, 193\AA\ and 211\AA\ data (in the second from left to the fourth panel, respectively). The color density reflects the number of the detected ARMC events.}
\label{groupsall}
\end{figure*}

An additional illustration of the considered grouping of the observed ARMCs in terms of their speed and intensity is given in the bivariate maps in Fig.~\ref{groupsall}. The second, third, and fourth panels of Fig.~\ref{groupsall} correspond to the SDO/AIA wavelengths 171\AA, 193\AA\ and 211\AA,\ respectively.

\subsection{Characteristics and grouping of ARMCs, according to Hi-C2.1 data}

\begin{figure}[ht!]
\centering
\includegraphics[width=80mm]{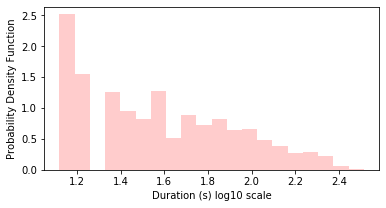}
\includegraphics[width=80mm]{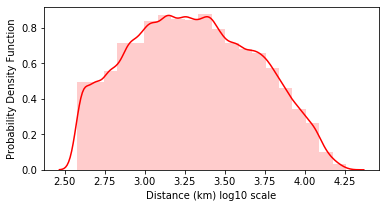}
\includegraphics[width=80mm]{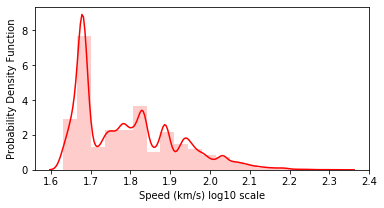}
\includegraphics[width=80mm]{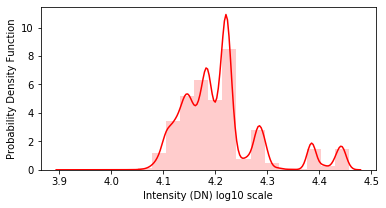}
\caption{Distribution of the ARMC characteristics in histograms (shaded vertical bars) and kernel densities (solid lines), revealed from the Hi-2.1 images. From top to bottom: Duration, distance, speed, and intensity. The duration and distance distributions are both unimodal, which indicates that they all belong to one group. The speed and intensity distributions are multimodal, which indicates the existence of ARMC groups.}
\label{histhic2}
\end{figure}

The distributions of duration, distance, speed, and intensity of the ARMCs observed in the Hi-C2.1 data are illustrated in Fig.~\ref{histhic2}. Similar to the case of the SDO/AIA flows, the distance and duration distributions are unimodal, which implies that only one group of ARMCs exists.\ On the other hand, the speed and intensity distributions are multimodal, indicating the presence of multiple groups per characteristic. 

\begin{figure}[ht!]
\centering
\includegraphics[width=80mm]{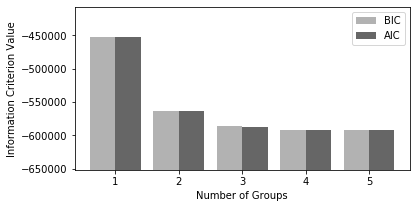}
\includegraphics[width=80mm]{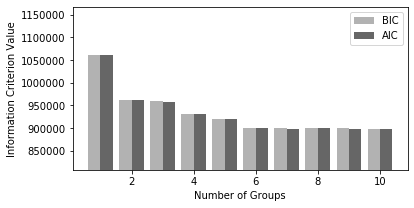}
\caption{BIC and AIC of the GMM fit to the Hi-C2.1 flow speed (top) and intensity (bottom).}
\label{hic2_bic_aic}
\end{figure}

To determine the number of groups, we again considered the BIC and AIC with the aim of striking a balance between a good distribution fit (the more groups, the better the fit) and the fact that the groups must be distinguishable (the fewer groups, the more they are distinguishable). Figure~\ref{hic2_bic_aic} shows that both the BIC and AIC values saturate at four and six groups for the flow speeds and intensities, respectively. This means that for the Hi-C2.1 data these values, similar to the SDO/AIA case, both have an adequate goodness-of-fit and are clearly distinguishable. 

The resulting GMM's goodness-of-fit was again assessed by the Anderson-Darling and Cramer-von Mises hypothesis tests. The results of these hypothesis tests are summarized in Table~\ref{speedtestshic} for the speeds and in Table~\ref{intensitytestshic} for the intensities.  

\begin{center}
\begin{table}[ht]
\begin{tabular}{c|c|c}
\hline \hline
    & Anderson-Darling & Cramer-von Mises \\
Number of groups        &  test statistic p-value &  test statistic p-value \\
\hline \hline 
1   & 0.00060  & $<$0.00001 \\
2   & 0.00060  & $<$0.00001  \\
3   & 0.00060   & 0.00033\\
4   & 0.02509   & 0.02940 \\
5   & 0.08274   & 0.07332 \\
\hline \hline
\end{tabular}
\caption{Statistical hypothesis tests for the GMM goodness-of-fit regarding the speeds of ARMCs observed with Hi-C2.1}
\label{speedtestshic}
\end{table}
\end{center} 

The p-values corresponding to the hypothesis tests that the observed flow speeds are generated from a GMM consisting one, two, or three groups are all less than 0.01\%. Therefore the hypothesis that the observed flow speeds are generated from a Gaussian distribution with three or fewer groups is rejected at a 99.99\% level of significance. The p-value corresponding to the hypothesis tests that the observed flow speeds are generated from a GMM of four groups is greater than 2.5\%. We therefore cannot reject the hypothesis that the observed flow speeds are generated from a GMM of four groups at a 97.5\% level of significance.  \\

\begin{center}
\begin{table}[ht]
\begin{tabular}{c|c|c}
\hline \hline
    & Anderson-Darling & Cramer-von Mises \\
Number of groups        &  test statistic p-value &  test statistic p-value \\
\hline \hline 
1   & 0.00060   & $<$0.00001 \\
2   & 0.03770   & 0.04027\\
3   & 0.09966   & 0.08369\\
4   & 0.15124   & 0.17106\\
5   & 0.19909   & 0.17752 \\
6   & 0.19612   & 0.17754 \\
\hline \hline
\end{tabular}
\caption{Statistical hypothesis tests for the GMM goodness-of-fit regarding the intensities of ARMCs observed with Hi-C2.1}
\label{intensitytestshic}
\end{table}
\end{center} 

\begin{figure*}
\centering
\includegraphics[width=80mm]{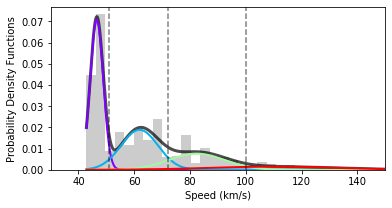}
\includegraphics[width=80mm]{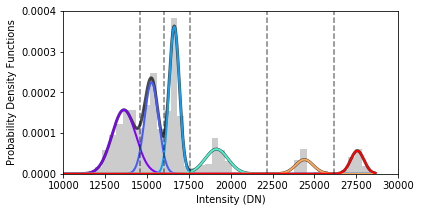}
\caption{GMM fit to the Hi-C2.1 flow speed (top) and intensity (bottom). The vertical lines represent the points where one Gaussian density crosses another, which indicates a separation of the ARMC groups. }
\label{hic2_gaussian_mix}
\end{figure*}

The p-values corresponding to the hypothesis tests that the observed flow intensities are generated from a GMM consisting one, two, or three groups are all less than 10\%. Therefore, we can reject the hypothesis that the observed flow intensities are generated from a Gaussian distribution with three or fewer groups at a 90\% level of significance. 
The p-value corresponding to the hypothesis tests that the observed flow intensities are generated from a GMM of four, five, or six groups is greater than 15\%. Therefore, we can not reject the hypothesis that the observed flow intensities are generated from a GMM of four, five, or six groups at a 85\% level of significance.  
Similar to the case of SDO/AIA data, the resulting parameter estimates for the ARMCs, as observed with Hi-C2.1, are used to group them according to their respective responsibilities. Figure~\ref{hic2_gaussian_mix} illustrates the Gaussian density functions of each ARMC group (colored lines) together with the resulting GMM's density function (black line). The group splits, at which a flow's responsibility to both groups is equal, are also shown. The ARMC speed and intensity grouping, as observed with Hi-C2.1, can also be seen in the bivariate map in Fig.~\ref{groupsall}, in the leftmost panel.

\begin{table*}
\caption{Distribution percentage of the discovered ARMCs over the groups according their speed and intensity. The color coding of groups is the same as in Figs.~\ref{gaussian_mix} and \ref{hic2_gaussian_mix}. The group numbering reflects the change in the corresponding grouping characteristic parameter (i.e., speed or intensity, specified in Table~\ref{intensitytestshicHIC}), i.e.,\ group~1 contains the events with the smallest magnitude of the grouping parameter, which then increases with increasing group number.}
\label{intensitytestshicAIA}
\centering 
\begin{tabular}{c c c c c c }
\hline \hline
Group & Characteristic  & Hi-C2.1   & SDO/AIA 171   & SDO/AIA 193   & SDO/AIA 211 \\ 
\hline  
\textcolor{violet}{1} & Speed  & 55 797 (37.6\%)  & 7 852 (27.5\%)   & 7 872 (22.7\%)   & 8 294 (26.3\%)  \\
\textcolor{blue}{2} & Speed & 50 518 (34.1\%)   & 9 334 (32.7\%)   & 10 833 (31.2\%)   & 11 670 (37.0\%)  \\
\textcolor{green}{3} & Speed    & 31 469 (21.2\%)  & 6 855 (24.1\%)   & 11 849 (34.1\%)   & 7 398 (23.5\%)  \\
\textcolor{red}{4} & Speed     & 10 560 (7.1\%)   & 4 481 (15.7\%)   & 4 184 (12.0\%)   & 4 168 (13.2\%)  \\
\textcolor{violet}{1} & Intensity    & 41 045 (27.7\%)  & 0 (0.0\%)    & 0 (0.00\%)    & 16 870 (53.5\%)  \\
\textcolor{blue}{2} & Intensity   & 34 164 (23.0\%)  & 5 346 (18.8\%)   & 15 796 (45.5\%)   & 11 221 (35.6\%)  \\
\textcolor{cyan}{3} & Intensity   & 43 925 (29.6\%)  & 13 129 (46.0\%)   & 15 266 (44.9\%)   & 2 832 (9.0\%)  \\
\textcolor{green}{4} & Intensity    & 16 029 (10.8\%)  & 7 623 (26.7\%)   & 1 319 (3.8\%)    & 330 (1.0\%)  \\
\textcolor{orange}{5} & Intensity    & 6 914 (4.7\%)    & 1 261 (4.4\%)    & 1 765 (5.1\%)    & 277 (0.9\%)  \\
\textcolor{red}{6} & Intensity    & 6 267 (4.2\%)    & 1 163 (4.1\%)    & 5 92 (1.7\%)   & 0 (0.0\%)  \\
\hline 
\end{tabular}
\end{table*}

\begin{table*}
\caption{Physical parameters of the respective speed and intensity groups of the discovered ARMCs. The color coding of groups is the same as in Figs. \ref{gaussian_mix} and \ref{hic2_gaussian_mix}. The groups are numbered according to the characteristic parameter mean value, with 1 the lowest.}
\label{intensitytestshicHIC}
\centering 
\begin{tabular}{c c|c c c|c c c}
\hline \hline
& & \multicolumn{3}{c|}{Hi-C2.1 Parameters} & \multicolumn{3}{c}{SDO/AIA Parameters} \\ 
Group & Characteristic   & Mean  & Variance  & Weight    & Mean  & Variance     & Weight \\ 
\hline  
\textcolor{violet}{1} & Speed &  47 km/s & 1 km/s & 0.37 & 39 km/s & 3 km/s & 0.25 \\
\textcolor{blue}{2} &Speed & 61 km/s & 6 km/s & 0.33 & 51 km/s & 3 km/s & 0.33 \\
\textcolor{green}{3} & Speed & 81 km/s & 10 km/s & 0.21 & 71 km/s & 7 km/s & 0.26 \\
\textcolor{red}{4} & Speed & 108 km/s & 22 km/s & 0.09 & 101 km/s & 22 km/s & 0.16 \\
\textcolor{violet}{1} & Intensity & 13579 DN &  698 DN & 0.28& 852 DN & 139 DN &    0.18\\
\textcolor{blue}{2} & Intensity & 15199 DN &    443 DN & 0.23 & 1587 DN & 222     DN & 0.34 \\
\textcolor{cyan}{3} & Intensity & 16619 DN &    317 DN & 0.29 & 2381 DN & 208 DN &        0.33\\
\textcolor{green}{4} & Intensity & 19148 DN & 660 DN & 0.11 & 3247 DN   & 218 DN  & 0.10 \\
\textcolor{orange}{5} &Intensity & 24403 DN & 543 DN & 0.05 & 4122 DN   & 126 DN & 0.03\\
\textcolor{red}{6} & Intensity & 27599 DN & 454 DN & 0.04 & 5114 DN & 246 DN & 0.02 \\
\hline 
\end{tabular}
\end{table*}

\subsection{Group locations}

The statistical analysis of the observations by SDO/AIA (in 171\AA, 193\AA\, and 211\AA) and Hi-C2.1 (in 172\AA) of the moving ARMC yields a grouping of characteristics, specifically in terms of intensity and speed (see Figs.~\ref{hist}, \ref{density}, \ref{histhic2}, and \ref{groupsall}). This grouping was also reproduced with the corresponding GMM distributions (see Figs. \ref{gaussian_mix} and \ref{hic2_gaussian_mix})

The main characteristics of the discovered ARMC groups and the related physical parameters of the observed flows are summarized in Tables~\ref{intensitytestshicAIA} and \ref{intensitytestshicHIC}. Specifically, in the data of both missions we can distinguish four groups of ARMCs with respect to their speed and six groups with respect to their EUV intensity. The range of the typical speeds of the observed flows appears to be from 40 km/s to 140 km/s.

\begin{figure*}
    \includegraphics[width=.24\textwidth]{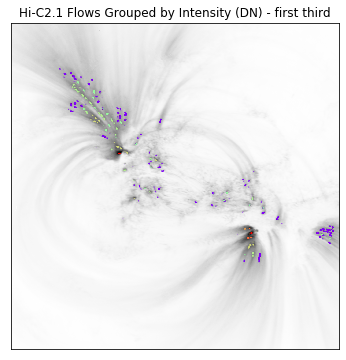}\hfill
    \includegraphics[width=.24\textwidth]{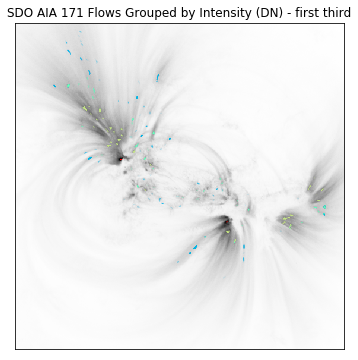}\hfill
    \includegraphics[width=.24\textwidth]{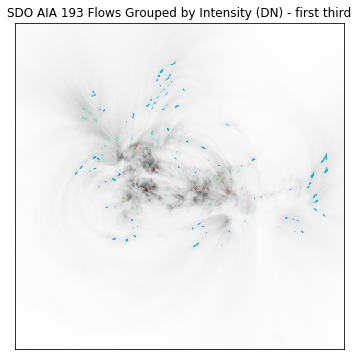}\hfill
    \includegraphics[width=.24\textwidth]{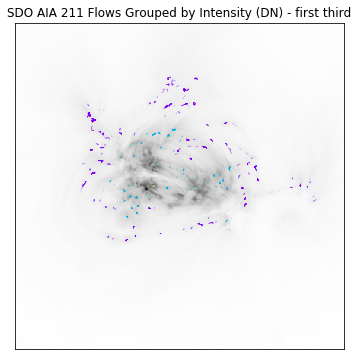}
    \\[\smallskipamount]
    \includegraphics[width=.24\textwidth]{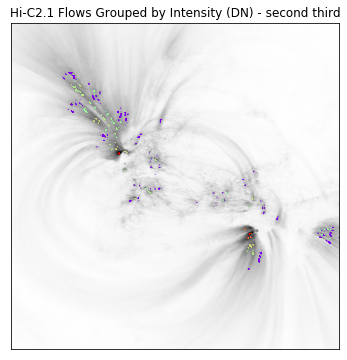}\hfill
    \includegraphics[width=.24\textwidth]{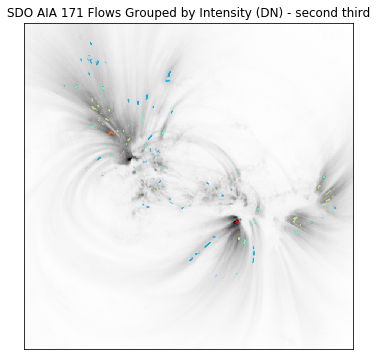}\hfill
    \includegraphics[width=.24\textwidth]{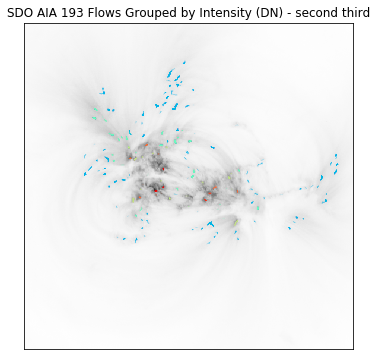}\hfill
    \includegraphics[width=.24\textwidth]{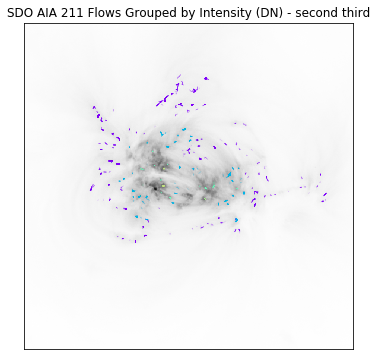}
    \\[\smallskipamount]
  \includegraphics[width=.24\textwidth]{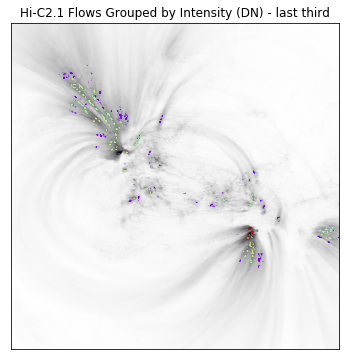}\hfill
    \includegraphics[width=.24\textwidth]{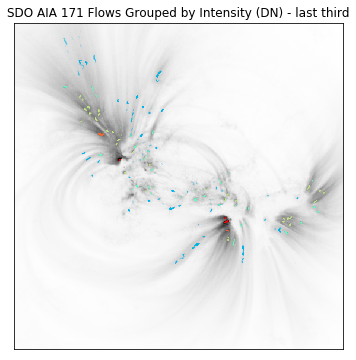}\hfill
    \includegraphics[width=.24\textwidth]{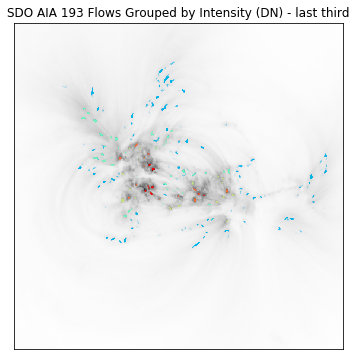}\hfill
    \includegraphics[width=.24\textwidth]{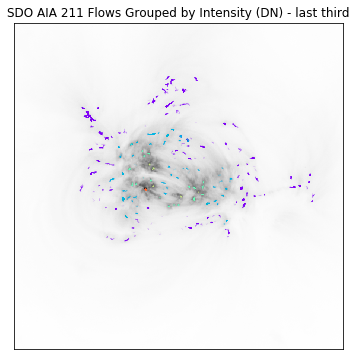}
\caption{Location of ARMCs in the active region observed during the consequent 111~s time intervals (different rows of panels), with the events' groupings with respect to the intensity indicated. The color coding of the groups is the same as in Figs. \ref{gaussian_mix} and \ref{hic2_gaussian_mix}. Columns from left to right represent observations with Hi-C2.1 (172\AA) and SDO/AIA 171\AA, 193\AA \ and 211\AA,\ respectively. The EUV images of the active region AR12712 loop structure were all taken on 29 May 2018 at 18:56:21.}
\label{global1}
\end{figure*} 

\begin{figure*}[ht!]
    \includegraphics[width=.24\textwidth]{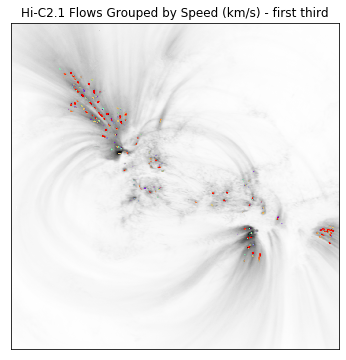}\hfill
    \includegraphics[width=.24\textwidth]{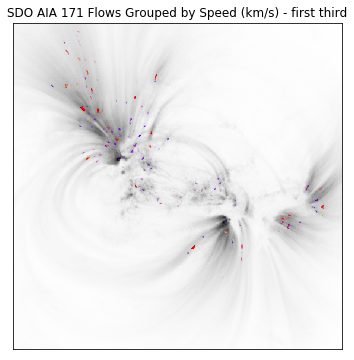}\hfill
    \includegraphics[width=.24\textwidth]{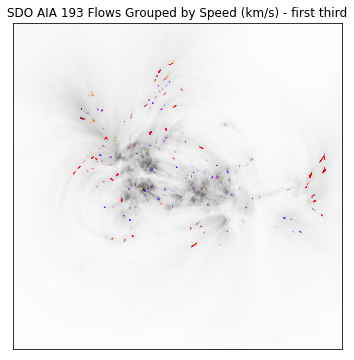}\hfill
    \includegraphics[width=.24\textwidth]{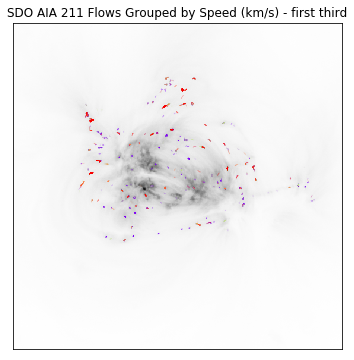}
    \\[\smallskipamount]
    \includegraphics[width=.24\textwidth]{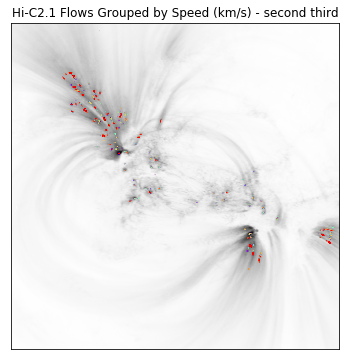}\hfill
    \includegraphics[width=.24\textwidth]{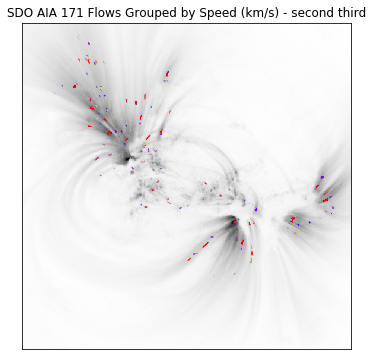}\hfill
    \includegraphics[width=.24\textwidth]{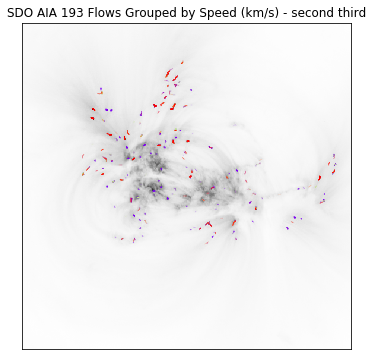}\hfill
    \includegraphics[width=.24\textwidth]{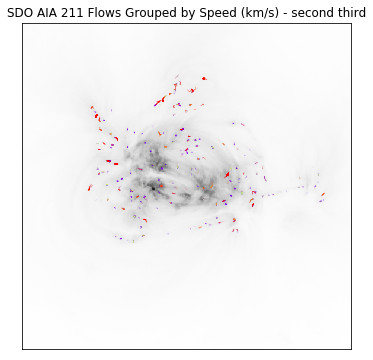}
    \\[\smallskipamount]
  \includegraphics[width=.24\textwidth]{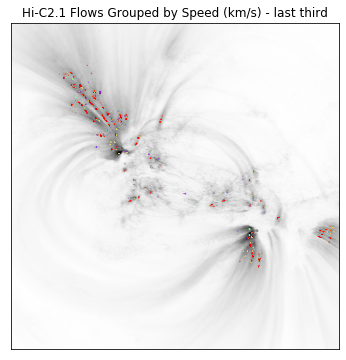}\hfill
    \includegraphics[width=.24\textwidth]{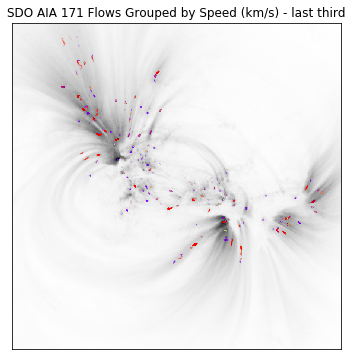}\hfill
    \includegraphics[width=.24\textwidth]{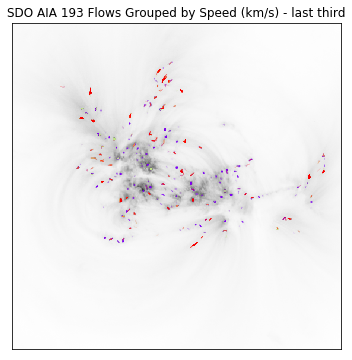}\hfill
    \includegraphics[width=.24\textwidth]{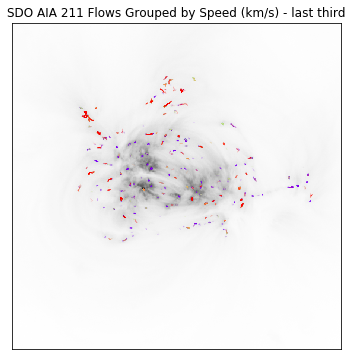}
\caption{Same as Fig. \ref{global1}, but for the ARMCs grouped with respect to speed.}
\label{global2}
\end{figure*} 
Figures \ref{global1} and \ref{global2} illustrate the ARMC locations in the active region's magnetic loop structure, indicating their grouping according to the intensity and speed, respectively. These figures additionally reveal the properties of the spatial distribution of the ARMC flows. Firstly, the intensity groups are clearly distinguishable in the images. The high intensity groups are concentrated near the spots, close to the loop foot points, whereas the lower intensity groups appear predominantly in the loop's periphery. This pattern remains consistent for both the Hi-C2.1 wavelength and all SDO/AIA wavelengths. Secondly, the distribution of the speed groups is less regular, which indicates that the observed ARMC speeds could be quite different and that they are evenly distributed over the various locations. This implies that many loops contain flows of varying speeds.
\section{Discussion}
The purpose of our study was to examine the hypothesis of whether the observed moving small-scale intensity enhancements in the solar coronal active regions, referred to as ARMCs, show any systematic spatiotemporal features and/or dynamic behavior. In order to achieve this goal, we applied a statistical inference methodology to the physical characteristics of ARMCs observed with SDO/AIA and Hi-C2.1. The analysis reveals a large number of ARMCs moving in the multistrand magnetic structure of the active region. The examination of data from two different missions with distinct spatiotemporal resolutions of EUV imagers enables the obtained results to be cross-validated and related conclusions to be drawn. Such an approach is invaluable for understanding the core configurations and dynamics of the solar coronal magnetic features \citep[as was addressed in, e.g.,][]{WilliamsHIC22020}. The general outcome of our comparative study vividly reveals that there are several groups of ARMCs with respect to intensity and speed visible in the SDO/AIA wavelengths 171\AA, 193\AA \ and 211\AA.\ These groups cannot be ascribed to any observational or instrumental artifact, as similar results were also obtained with Hi-C2.1 data with completely distinct  technological characteristics. Therefore, the observations qualitatively confirm (as shown in Sect.~2) that the observed grouping of ARMCs along their trajectories is a physical phenomenon. With regard to the quantitative differences, they are related to the instrumental features (such as charge-coupled device sensitivity, its resolution, time cadence, intensity unit, etc.), which do not affect the main conclusions drawn from our analysis. 
It is evident from the applied statistical methodology that we are dealing with different groups of flows manifested in the considered  SDO/AIA and Hi-C2.1 wavelengths, 171\AA, 172\AA, 193\AA, \ and 211\AA \ as characterized in terms of the local maxima  of the probability densities with respect to the 
speed and intensity (see Figs. \ref{hist} and \ref{histhic2}, respectively, as well as the upper and lower parts of Table~\ref{intensitytestshicHIC}).

The probability densities of intensities (illustrated in Fig.~\ref{hist} and summarized in  Table~\ref{intensitytests}) at rather high DN rates range from $2.7-3.7$ powers of 10 (with an evidently multimodal distribution, as is argued throughout the paper), while the rest of the wavelengths show significantly lower characteristic intensities (DN), ranging from $1.0-2.5$ powers of 10 (the presence of the multimodal structure of the distribution in this case is doubtful, albeit not excluded).
It is noticeable that, regardless of the physical nature of the observed ARMCs,
in all wavelengths their characteristic velocities fall within approximately the same range of $40-140\;$km/s (top panel in Fig.~\ref{density} and upper part of the Table~\ref{intensitytestshicHIC}). 

{It should be noted here that during the evaluation of the centroid motion speed, the projection effect related to the spherical shape of the Sun is removed using the algorithm described in \citet{Philishvili2021}, which implies a corresponding resizing of the pixels depending on their position with respect to\ the visible center of the solar disk in the considered EUV images. The projection effect, related to the 3D magnetic loop structure and orientation of the line-of-sight (LOS), is only partially taken into account, by setting the threshold of the velocity. In fact, only the flows above the adopted threshold are taken into consideration. The magnetic field orientation effect can only underestimate the moving velocity, and in a critical case, when the blob moves along the LOS, it seems to be a standing point in the 2D  EUV images. Therefore, by setting a lower limit for the speed, we guarantee that the flows included in our statistics either move strictly perpendicular to the LOS (most probably) or in a direction slightly inclined to the LOS. Accordingly, the projection-effect-related underestimation of the velocities in the latter cases is limited.}

Altogether, for all the SDO/AIA wavelengths, 171\AA, 193\AA\ and 211\AA,\ we find four main groups with respect to the speed distributions and six groups for the observed intensities (Figs.~\ref{gaussian_mix} and \ref{groupsall}). In the Hi-C2.1 172\AA data, again six groups for the intensities and four groups for the speed (Figs.~\ref{hic2_gaussian_mix} and \ref{groupsall} (left panel)) have been found with high levels of statistical confidence. 

It is worth mentioning, however, that for both missions the four groups with respect to the speed are identified with a considerably smaller significance level (see, e.g., Figs. \ref{gaussian_mix} and \ref{hic2_gaussian_mix}). It is also important to notice that, according to Fig.~\ref{groupsall}, the grouping of ARMCs in the SDO/AIA wavelength 171\AA \ has a peak in the intensity range of 2000-3000 DN, whereas in the 193\AA\  and 211\AA\ wavelengths the maxima of all groups shift toward the lower intensity values. At the same time, the group size with respect to the speed in the whole range of $40-140\;$km/s does not change significantly when viewed in different wavelengths.

The main results of our study are visualized in Figs.~\ref{global1} and \ref{global2}, where the spatial locations of the ARMC groups in terms of intensity and speed are shown separately. They illustrate the spatial distribution (concentration) maps of the identified moving ARMC features within the strand structure of the active region AR 12712. The three rows in these figures correspond to three different time intervals of 111 seconds each, specifically: 18:56:21 -- 18:58:12 UT (top row), 18:58:13 -- 19:00:04 UT (middle row), and 19:00:05 -- 19:01:56 UT (bottom row). The total observational time of the Hi-C2.1 mission is also divided into the same three equal consecutive subintervals. A visual inspection of all panels for the intensities (Fig.~\ref{global1}) and speeds (Fig.~\ref{global2}) indicates that the ARMCs follow the local main structure (plausibly the force-free part) of the active region's magnetic field. The maps of the intensity enhancements for Hi-C2.1 172\AA \ (left panels in all rows) and SDO/AIA 171\AA \ (second from the left panels in all rows) are similar, which is expected. In particular, it is seen that the high intensity groups (yellow and orange) are predominantly located close to the {areas of the concentrated magnetic flux} (loop foot points), while the intermediate and low intensity groups are spread along the loops. For the Hi-C2.1 and SDO/AIA 171\AA \ flows, the intensity of ARMCs decreases with increasing distance 
from the foot points. The picture is similar for the ARMCs in 193\AA \ (third panel from the left in all rows) and in 211\AA \ (right panels in all rows). However, the  medium and low intensity patterns, observed in the 193\AA \ flows (mostly cyan) and the 211\AA \ flows (mostly violet), are located farther away from the foot points and predominantly cover the magnetic loop apex areas.  
The situation is somewhat altered for the location of ARMC groups with respect to the speed (Fig.~\ref{global2}). In the case of the Hi-C2.1 172\AA \ and SDO/AIA 171\AA \ data, the intermediate and low velocity trajectories are primarily concentrated closer to the foot point regions than the higher-velocity groups, although all these patterns are complemented by intermediate (and even lower) velocity events. The picture is similar for the 193\AA \ and 211\AA \ data accordingly, with the specifics analogous to the case of intensity maps discussed above (Fig.~\ref{global1}).  
\section{Conclusions}
The characteristic velocities of the considered ARMCs are almost independent of the wavelength or intensity rates.\ This allows us to conclude that we are dealing with the manifestation of the same physical mechanism responsible for the acceleration of ARMCs at different altitudes monitored in the corresponding wavelengths. It is reasonable to suppose that the acceleration depends rather on the pressure difference between the start and the end of the flow trajectories, predominantly along the force-free or even potential magnetic field configurations of the active region magnetic loops; however, the occurrence of the flow patterns {related} to the non-potential or non-force-free components of the magnetic field is not excluded. We find evidence that the statistically inferred number of groups for the intensities is six and for the speeds, four. This may be counted as an argument that the physical processes responsible for the grouping with respect to intensity could differ from those resulting in the speed grouping and, therefore, the ARMC acceleration.

In particular, the observed local brightenings related to the increased temperature can be attributed to the statistics of the magnetic reconnection events. The ARMC motion with the characteristic velocity around the sound speeds or below can be due to a siphon-type acceleration that is related to the pressure gradients appearing along the loop strands \citep{montesinos1993}. These are small-scale counterparts of the usual large-scale solar wind patterns \citep[e.g., see][and references therein]{Shergelashvili-etal-2020} or the flows in a thermally nonequilibrium medium \citep{Shergelashvili2007}. A definite conclusion from our study is that the ARMCs are events that are fundamentally different than the blobs that have a quasi-periodic behavior in the flaring active regions analyzed in \citet{Philishvili2017}. 

The velocity groups are concentrated within the range $40-140\;$km/s. This is a natural result, as shown in \citet{Philishvili2021}. {According to our estimations based on the dynamic emission measure analysis}, the local background sound speed appears to be in the range of $100-120\;$km/s. {However, the sound speed is not only determined by the temperature estimation: it is also affected by the conditions of the local thermodynamic nonequilibrium or thermal imbalance \citep{Zavershinskii2019}. The} majority of identified ARMC flows are expected to be subsonic, and only a small fraction of them are supersonic. One more interesting consequence of the performed mapping of the ARMCs is that in all the considered wavelengths we also see the presence of consecutive low- and high-velocity ARMCs (``pairs''), which might imply that we are dealing with flow acceleration or deceleration patterns. A rigorous theoretical modeling is planned in the future to address this issue and to reveal the properties of such flows. For instance, we plan to develop an ARMC acceleration model similar to that in \citet{montesinos1993} and analogous to those used for the large-scale flow patterns in the Parker solar wind model or that which recently reported the model's discontinuous counterpart \citep{Shergelashvili-etal-2020}. 

The potential use of the maps we obtained in the paper enables an improvement in the magnetic field extrapolation algorithms through the employment of the moving ARMC intensity and speed pattern libraries in the various artificial intelligence applications. The results obtained in this paper can serve as a background for several further studies, such as the development of a statistical model of small-scale flow patterns that can represent effective ``fingerprints'' of active regions. Based on such statistical pattern modeling, in combination with corresponding analytical theories of flows, learning and testing sets of active region images can be collected for the corresponding machine and deep learning engines.

\begin{acknowledgements}
The work was developed under the support of Shota Rustaveli Georgian National Science Foundation grant FR17\_609 and the Ilia State University Institutional Development Grant "Classification of synoptic patterns of the inner heliosphere for the AI interdisciplinary models". These results were also obtained in the framework of the projects C14/19/089  (C1 project Internal Funds KU Leuven), G.0D07.19N  (FWO-Vlaanderen), SIDC Data Exploitation (ESA Prodex-12), and Belspo projects BR/165/A2/CCSOM and B2/191/P1/SWiM. M.L.K. additionally acknowledges the project S11606-N16 of the Austrian Science Fund (FWF) and grant No. 075-15-2020-780 (GA No. 13.1902.21.0039) of the Russian Ministry of Education.
\end{acknowledgements}

\bibliographystyle{aa}
\bibliography{biblio}

\end{document}